\def\Roman#1{\uppercase\expandafter{\romannumeral#1}}
\documentclass[a4paper,12pt]{article}
\usepackage{cite}
\usepackage{amssymb,amsfonts}
\usepackage[mathscr]{eucal}
\usepackage{mathrsfs}

\title{Path integral on a manifold with a non-free group action }

\author{S. N. Storchak\footnote{E-mail adress: storchak@ihep.ru}\\
\small{Institute for High Energy Physics, Protvino, Moscow Region, 142284, Russia}}

\begin{document}

\maketitle

\begin{abstract}
The method of the factorization of the path integral measure, based on a nonlinear filtering equation, is extended to the case of a nonfree isometric action of the compact semisimple unimodular Lie group on a smooth compact Riemannian manifold. The method is applied to the path integral which describes the  "quantum'' motion of  the scalar particle on this manifold. The relation between path integral representing the solution of the parabolic equation on initial and reduced manifold is derived. It is shown that reduction  reduction leads to the non invariance of the path integral measure.  

\end{abstract}
{\bf{keywords:}} Marsden-Weinstein reduction, Kaluza--Klein theories, path integral, stochastic analysis.\\

\section*{Introduction}
In many cases, the presence of symmetry in dynamical systems allows us to find an exact solution of the corresponding evolution problem.
This is because the original dynamical system can be reduced to another system with fewer degrees of freedom.
There  is a rather important class of such systems, which describe the motion of scalar particles on manifolds with a given group action. 
If the action of a group is free then the original manifold can be viewed as a total space of the principal fiber bundle.

A more complicated case of a non-free group action 
on a manifold was also studied in many published papers \cite{Borel, Janich, Bredon, Davis, Palais, Duist, Pflaum}. 
In this case, the points of the original manifold may have the different isotropy subgroups.  Therefore the orbits may  also be quite different. 
This brings us to the classifications of 
the points of the original manifold based on their isotropy subgroups. 
The points with the conjugated isotropy subgroups (by an inner automorphism of the group) form a class of points of  a definite  orbit type (or stratum). 
The original manifold becomes a collection of these strata, 
and for this reason, the orbit space is also stratified. 

The basic facts on the  classical dynamical systems given on stratified  ``manifolds'' and their reductions can  be find in \cite{Lerman_1, Lerman_2, Ortega}. Much less is known about  quantization of these systems. 
In this regard, we mention the articles \cite{Iwai, Tanimura} and \cite{Landsman} where the  questions related to the quantization of  such  systems were considered. 
  
But in general, the quantization problem of these systems is far from a final solution. 
 However,  since we are meeting with a non-free group action  
as in the gauge field theory \cite{Rudolph, Fleischhack_1, Fleischhack_2} and in various finite-dimensional dynamical systems with symmetry,
the need for its solution remains actual.
In the present paper we confine ourselves to the consideration of the particular case. Namely, we assume that we deal with   a smooth proper non-free isometric action of the compact semisimple (unimodular) Lie group on the original compact Riemannian manifold. It is also assumed that as a result of the action 
there is only a single stratum (i.e., all  points of manifold belong to a single orbit type).

Precisely such a case   of a group action arises in the study of multi-dimensional field theories which assume that the inner and outer space are not independent but are manifestations of a single manifold of higher dimension \cite{Coquereaux, Jadczyk, Jadczyk_book, Cho}.
This case also occurs in the study of various approaches to the symplectic reduction in mechanics \cite{Hochgerner_1, Hochgerner_2}

In the present paper, the dynamical system under the study is a system given by a scalar particle motion on a manifold. 
The ``quantum'' evolution of the system is studied with the help of the path integration. 

We use the definition of our path integrals from \cite{Daletskii}. These integrals represent  solutions of the backward Kolmogorov equations given on a manifold. By definition, the path integral measure is generated by the stochastic process (also given on a manifold). The process, in turn, is determined by the solution of a stochastic differential equation. The transformation of the path integral measure is related to the corresponding transformation of the stochastic differential equation which governs the local evolution.

Due to the presence of symmetry in our dynamical system (provided  we have a group invariant Hamiltonian),  an initional system is reduced to another dynamical system defined on the orbit space. We will study the reduction of the original path integral and its relation with the path integral on the orbit space.
In case of a free action of a group on a manifold, we have used \cite{Storchak} the nonlinear filtering equation from the stochastic process theory  in order to factorize  the path integral measure, thereby   reaching the separation of   variables in the path integral associated with the group.
Later,  a similar approach to the factorization of the measure has been used   in \cite{Elworthy}. 

In the present paper, it will be shown that the case of a non-free group action, leading to a single orbit type,  can be examined in a similar way.  The relation between the original path integral and the path integral on the orbit space will be obtained by making use of the special transformation of the path integral measure. This transformation is also derived with the help of the nonlinear filtering equation.

In  Section 1 of the paper  we give a short introduction to the geometry that arises in the problems of reduction of dynamical systems given on manifolds in case of a  non free action of a Lie group.  

In  Section 2, we discuss questions related with the definition of the Riemannian metric with  the help of the coordinates of the associated bundle.

Section 3 considers the definition of the path integrals based on the local stochastic processes and the questions of the transformation of these integrals. 

In Section 4,  we derive the nonlinear filtering equation for the stochastic processe that are used in our problem and then study the transformation of the path integral  which leads to the factorization of the measure. Also in this section we obtain the relation between the path integral given on the original manifold and the integral on the orbit space of the group action.

\section{The original manifold  as a bundle space}
We are dealing with  a proper non-free isometric action of a compact (connected) semisimple  group Lie $\mathcal G$ on a smooth compact manifold $\mathcal E$. We  consider the case when  the  group $\mathcal G$ acts on $\mathcal E$ on the right. Our main assumption is concerned  the orbits of this action. We suppose that all orbits of $\mathcal G$ belongs to a single orbit type. This means that for each $q \in \mathcal E$, its isotropy subgroup $\mathcal G_q$ is congugated to some  standard subgroup $\mathcal H$ within $\mathcal G$: $\mathcal G_q=g\mathcal H g^{-1}$. Thus, in our case $\mathcal E=\mathcal E_{(\mathcal H)}=\{q\in \mathcal E : \mathcal G_q=g\mathcal H g^{-1}\}$.

As a result, all the orbits are isomorphic to the factor space $\mathcal G \setminus \mathcal H$  consisting of the right classes $\mathcal H g$, and the manifold $\mathcal E$ is bundle space having $\mathcal M=\mathcal E / \mathcal G$ as a base. So $\mathcal E$ can be locally presented as $\mathcal M \times \,\mathcal G \backslash \mathcal H$.

But $\mathcal E$ is not a principal fiber bundle. It was shown in \cite{Borel, Marsden, Palais, Duist, Pflaum} that  $\mathcal E=\mathcal E_{(\mathcal H)}$ has a structure of an associated fiber bundle  $\hat{\mathcal E}=(\mathcal P {\times}_{\mathcal K}\,\, \mathcal G \backslash \mathcal H)$ over $\mathcal M$, where  the manifold $\mathcal P=\mathcal E_{\mathcal H}=\{q\in \mathcal E : \mathcal G_q=\mathcal H \}$, the factor group $\mathcal K =N(\mathcal H) / \mathcal H$ with the normalizer of $\mathcal H$ in $\mathcal G$ defined as   $N(\mathcal H)=\{g\in \mathcal G : g\mathcal H g^{-1}=\mathcal H \}$.
Note also that  right and left cosets in $\mathcal K$ coincide. 

The principal  bundle $P(\mathcal M, \mathcal K)$ with  total space $\mathcal P$ is associated to the bundle {$\hat{ \mathcal E}$}.
$\mathcal P$ is a submanifold in the original manifold $\mathcal E_{(\mathcal H)}$, and the group $\mathcal K$ acts freely on $\mathcal P$. 
The correspondence between the original manifold $\mathcal E_{(\mathcal H)}$ and the associated bundle $\hat {\mathcal E}$ is achieved in the following way.

To each element $[p,Hg]\equiv [p,[g]]$ \footnote{$[p,Hg]$ is an equivalence class of the elements $(p,Hg)\in \mathcal P{\times}\mathcal G \backslash \mathcal H$ with respect to the following equivalence relation: $(pk,k^{-1}Hg)\sim (p,Hg)$, $k\in \mathcal K$.}
in the associated bundle $\hat{\mathcal E}=(\mathcal P {\times}_{\mathcal K}\,\, \mathcal G \backslash \mathcal H)$ there corresponds the element $pg$ belonging to $\mathcal E_{(\mathcal H)}$. 
 On the other hand, if $q\in \mathcal E_{(\mathcal H)}$, then  $\mathcal G_q=g_{0}\mathcal H g^{-1}_{0}$ with some $g_{0} \in \mathcal G$. That is, 
$q(g_{0}\mathcal H g^{-1}_{0})=q$, or $qg_{0}\mathcal H =qg_{0}$. It means that $p=qg_{0}\in \mathcal P$. 
Taking into account the $(\bmod\,{\mathcal K})$-ambiguity of the solution $g_{0}$, we come to the correspondence between $\mathcal E_{(\mathcal H)}$ and $(\mathcal P {\times}_{\mathcal K}\,\, \mathcal G \backslash \mathcal H)$: $q\to [qg_{0},Hg^{-1}_{0}]$.

We may think about this correspondence as on the coordinatization of ${\mathcal E}_{(\mathcal H)}$.
It follows that in studying  the evolution given on the manifold $\mathcal E_{(\mathcal H)}$, one can use the appropriate coordinates of the  associated bundle $\hat{\mathscr E}$. 

In the sequel we will be  interested in the action of the original group (and  of its subgroups) on the manifold $\mathcal E_{(\mathcal H)}$. In coordinates of the associated bundle, the global (right) action of $\mathcal G$ is determined in the following way: $g': [p,[g]]\to [p,[gg']]$. 
As for  the  action (from the left) of the structure group $\mathcal K$ on a fiber of the associated bundle, we only note that this action commutes with the right action of $\mathcal G$.

Note also that these matters, and many others that are  related to the problem under the consideration, were discussed in detail in \cite{Coquereaux, Jadczyk, Jadczyk_2, Jadczyk_book}. We refer to these works for further information.

Before proceeding to the   definition of the coordinate expression of the  Riemannian metric on the manifold $\hat{\mathscr E}$, we consider the usual choice \cite{Coquereaux} of the basis in  the algebra Lie $\mathscr G$ of the group $\mathcal G$. This choice is related to the reductive decomposition of the algebras $\mathscr G$ and $\mathscr N$. 

For the algebra $\mathscr G$, we have
\[
 \mathscr G=\mathscr H+{\mathscr S}_{\mathcal H},\;\;\;\;\;({\rm Ad\, \mathcal H})({\mathscr S}_{\mathcal H})={\mathscr S}_{\mathcal H},
\]
where ${\mathscr S}_{\mathcal H}$ is the orthogonal complement of $\mathscr H$ in $\mathscr G$. It can be  identified with the vector  space
 tangent to homogeneous space $\mathcal G \backslash \mathcal H$ at the origin.

And for the Lie algebra $\mathscr N$ of the group $N(\mathcal H)$:
\[
 \mathscr N=\mathscr H+{\mathscr K}_{\mathcal H},\;\;\;\;\;({\rm Ad\, \mathcal H})({\mathscr K}_{\mathcal H})={\mathscr K}_{\mathcal H}.
\]
Note that ${\mathscr K}_{\mathcal H}=\mathscr N\cap{\mathscr S}_{\mathcal H}$, and ${\mathscr K}_{\mathcal H}$ is the orthogonal complement of ${\mathscr H}$ in ${\mathscr N}$. Also we have $[{\mathscr H},{\mathscr K}_{\mathcal H}]=0$. It means that ${\mathscr K}_{\mathcal H}$ is the subspace of the vectors in the tangent space ${\mathscr S}_{\mathcal H}$ that are invariant under the action of the group $\mathcal H$.

Besides, since $[{\mathscr K}_{\mathcal H},{\mathscr K}_{\mathcal H}]\subset{\mathscr K}_{\mathcal H}$, ${\mathscr K}_{\mathcal H}$ is identified with  the Lie algebra of the group $\mathcal K =N(\mathcal H) / \mathcal H$. So, ${\mathscr K}_{\mathcal H}$ can be also denoted by ${\mathscr K}$.

In a similar way we can consider the following decomposition:
\[
 \mathscr G=\mathscr N+{\mathscr L}_{\mathcal N},\;\;\;\;\;({\rm Ad\, \mathcal N})({\mathscr L}_{\mathcal N})={\mathscr L}_{\mathcal N}.
\]
Then 
\[
 {\mathscr S}_{\mathcal H}={\mathscr K}_{\mathcal H}+{\mathscr L}_{\mathcal N},\;\;\;\;\;({\rm Ad\, \mathcal H})({\mathscr L}_{\mathcal N})={\mathscr L}_{\mathcal N}.
\]
Thus we come to the following decomposition of $\mathscr G$: $$\mathscr G=\mathscr H\oplus{\mathscr K}_{\mathcal H}\oplus{\mathscr L}_{\mathcal N}$$ for which  
${\mathscr N}={\mathscr H}\oplus{\mathscr K}_{\mathcal H}$ and ${\mathscr S}_{\mathcal H}={\mathscr K}_{\mathcal H}\oplus{\mathscr L}_{\mathcal N}$.

The obtained   relations,
\begin{eqnarray*}
 &&[\mathscr H,\mathscr H]\subset\mathscr H, \;\;[{\mathscr K}_{\mathcal H},{\mathscr K}_{\mathcal H}]\subset{\mathscr K}_{\mathcal H},
\;\;[\mathscr H,{\mathscr K}_{\mathcal H}]=0,
\nonumber\\
&&[\mathscr H,{\mathscr L}_{\mathcal N}]\subset {\mathscr L}_{\mathcal N},\;\;
[{\mathscr K}_{\mathcal H},{\mathscr L}_{\mathcal N}]\subset{\mathscr L}_{\mathcal N},
\end{eqnarray*}
allows one to find the basis of the algebra ${\mathscr G}$ adapted to the above decomposition.

We will use the following convention about indices for  the generators  $Q_A$, $A=1,\ldots ,\dim {\mathcal G}$, of the Lie algebra $\mathscr G$:
\[
Q_A=\{Q_h,Q_a:\; Q_h\in \mathscr H,\, Q_a\in {\mathscr S}_{\mathcal H}\},  
\]
\[
 Q_a=\{Q_{\hat a},Q_{\bar a}:\;Q_{\hat a}\in {\mathscr K}_{\mathcal H},\, Q_{\bar a}\in {\mathscr L}_{\mathcal N}\}.
\]
In other words, the index $A=(h,a)$ and $a=(\hat a, \bar a)$, with 
$A=1,\ldots ,\dim {\mathcal G}$,  $h=1,\dots , \dim {\mathcal H}$, $a=1,\ldots,\dim {\mathcal G}\backslash {\mathcal H}$.

In these notations, the commutation relations are as follows:
\begin{eqnarray*}
&&[Q_{h_1},Q_{h_2}]=f^{\;\;\;\;\;h_3}_{h_1 h_2}Q_{h_3},\;\;\;
[Q_{h},Q_{a}]=f^{\;\;\;b}_{h a}Q_{b},
\nonumber\\
&&[Q_{a},Q_{b}]=f^{\;\;\;c}_{a b}Q_{c}+f_{a b}^{\;\;\;h}Q_h.
\end{eqnarray*}
 After the decomposition $Q_a=(Q_{\hat a},Q_{\bar a})$ they can be rewritten as
\begin{eqnarray*}
&&[Q_{\hat a},Q_{\hat b}]=f^{\;\;\;\hat c}_{\hat a \hat b}Q_{\hat c},\;\;
[Q_{h},Q_{\hat a}]=0,\;\;
[Q_{\hat a},Q_{\bar a}]=f^{\;\;\;\bar b}_{\hat a \bar a}Q_{\bar b},
\nonumber\\
&&[Q_{h},Q_{\bar a}]=f^{\;\;\;\bar b}_{h \bar a}Q_{\bar b},\;\;
[Q_{\bar a},Q_{\bar b}]=f^{\;\;\;\bar c}_{\bar a \bar b}Q_{\bar c}+f^{\;\;\;h}_{\bar a \bar b}Q_{h}.
\end{eqnarray*}

\section{The Riemannian metric }
In  reduction problems related to the classical mechanics (the Marsden-Weinstein reduction), the metric which is  given on the original manifold with a symmetry, determines  all necessary elements of the reduction process: the metric on the orbit space, the orbit metric and the mechanical connection in the corresponding principal bundle.
It  was shown (for example, in \cite{Coquereaux}) that the same is also true in our case, when the original manifold has a structure of the special associated bundle.

We have seen that   there is an  interrelation between the points of the original manifold $\mathcal E_{(\mathcal H)}$ and the points of the associated bundle $\hat{\mathscr E}$. This allows us to confine ourselves to consideration of  the ``quantum''
evolution ``given on $\hat{\mathscr E}\,$'',  since the evolution given on the original manifold is derivable from it.  To be precise,  our first object for study is the metric on ${\mathcal E}$ given in  coordinates of the associated bundle $\hat{\mathscr E}$. 

This metric should satisfy the definite requirements. The main of which is the right-invariance. Besides, since we deal with the bundle, the metric must be left-invariant with respect to the action of the structure group of the bundle, that is, with respect to the action of the group $\mathcal K =N(\mathcal H) / \mathcal H$. In some sense,  this approach to definition  of the metric (with such properties) is  an inverse procedure with respect to what is done in reduction. But we note that this lead to the same result. 

First of all, in order  to define the metric, we introduce a local coordinates on the associated bundle $\hat{\mathscr E}$. We will assume that there is some cross section $\sigma$ of the principal bundle $P(\mathcal M, \mathcal K)$ by which the point $p$ can be defined  as 
$p\sim(x,\sigma(x))$. On the (right) coset manifold  $\mathcal G \backslash \mathcal H$, the fiber of our associated  bundle $\hat{\mathscr E}$ at the point $x$ of the base,  we will use the coordinates $y^{\mu}$. So, in a local chart, the components of our metric depends on  $(x^i,y^{\mu})$. 

To take into account the invariant properties of the metric, we are need also to introduce group-valued coordinates on the  coset manifold 
$\mathcal G \backslash \mathcal H$. It is done with the help of the cross section $L_y$ of the principal bundle $\mathcal G \to \mathcal G \backslash \mathcal H$ \cite{Salam,Mecklenburg}. $L_y=e^{y^a Q_a}$ is a representative element of the right coset $\mathcal H g$. The right translation  of $L_y$ by $g$ is given by $L_yg=hL_{y'}$, where the element $h\in \mathcal H$  depends on $y$ and $g$. It follows that $L_{y'}=h^{-1}L_{y}g$. This transformation  is used for definition of the Killing vectors on $\mathcal G \backslash \mathcal H$.
 
In our case of the right action of the group $\mathcal G$, the Maurer-Cartan 1-form is defined by 
$$dL_y\, L^{-1}_y=\tilde{\rm e}^A(y)\,Q_A,\;\;\;\;\tilde{\rm e}^A(y)=\tilde{\rm e}^A_{\alpha}(y)\,dy^{\alpha}.$$
The right-invariant 1-form $\tilde{\rm e}^A(y)$ satisfies 
\[
 d\tilde{\rm e}^A=\frac12\,f^A_{BC}\,\tilde{\rm e}^B \wedge \tilde{\rm e}^C.
\]

Notice that $\tilde{\rm e}^A\,Q_A=\tilde{\rm e}^a\,Q_a+\tilde{\rm e}^h\,Q_h,$ in which $\tilde{\rm e}^a=
\tilde{\rm e}^a_{\mu}dy^{\mu}$ is known as a coframe. The vectors $\tilde{\rm e}^a_{\mu}$, $a=1,\ldots,\dim \mathcal G \backslash \mathcal H$,
form the set of the covariant basis vectors. It can be defined the reciprocal basis $\tilde{\rm e}_b^{\mu}$ such that 
$\tilde{\rm e}^a_{\mu}\tilde{\rm e}_b^{\mu}={\delta}_b^{\,a}$ and $\tilde{\rm e}^a_{\mu}\tilde{\rm e}_a^{\nu}={\delta}_{\mu}^{\,\nu}$.
 
In the paper, we adopt the following definition of the matrices $D^A_B$ of the adjoint representation of the group $\mathcal G$:
$$L_y\, Q_B\, L_y^{-1}=D^A_B(L_y)Q_A.$$
As  already noted,  transformation of $L_y$ leads to the Killing vectors on the coset manifold  $\mathcal G \backslash \mathcal H$. In our case, the Killing vectors for the right isometry of the group $\mathcal G$ are given by
$$K_A=K^{\alpha}_A(y) \frac{\partial}{\partial y^{\alpha}}\;\;\;\;\rm{with}\;\;\;\;
K^{\alpha}_A =D^a_A(L_y)\,\tilde{\rm e}^{\alpha}_a(y).$$
The Killing vectors have a standard commutation relation: $[K_A,K_B]=f^C_{AB}K_C$.
 The left isometry  on $\hat\mathscr E$ is realized by the Killing vectors of the group $N(\mathcal H)/\mathcal H$ which, in our basis, are given by $\tilde K^{\alpha}_{\,\,\hat b}=\tilde{\rm e}^{\alpha}_{\,\,\hat b}(y).$

It can be shown that in the local basis $(\frac{\partial}{\partial x^i}, \frac{\partial}{\partial y^{\alpha}})$, the metric which fulfill to our requirements can be written  in the following form:
\begin{equation}
\displaystyle
{ G}_{\mathcal A\mathcal B}=\left(
\begin{array}{cc}
h_{ij}+{\gamma }_{\alpha\,\beta}\, B_i^{\alpha}B_j^{\beta} & B_i^{\alpha}\, 
{\gamma }_{\alpha\,\beta} \\ {\gamma}_{\alpha\,\beta}\, B_i^{\beta}&{\gamma }_{\alpha\,\beta }
\end{array}
\right).
\label{a1}
\end{equation}
In this expression,   $h_{ij}(x)$   is the metric given on a manifold $\mathcal M$. The metric ${\gamma}_{\alpha\,\beta}$, 
$${\gamma}_{\alpha\,\beta}(x,y)=g_{a b}(x)\,\tilde{\rm e}^a_{\alpha}(y)\,\tilde{\rm e}^b_{\beta}(y),$$
is the metric on the fiber of the bundle $\hat\mathcal E$ taken at the point  $x$ of the base. 
$g_{a b}$ is not an arbitrary, but must be $\rm{Ad} \mathcal H$ invariant in order to  the metric  ${ G}_{\mathcal A\mathcal B}$ could be invariant under the aforementioned  actions of the groups $\mathcal G$ and $\mathcal K$.

The connection $B_i^{\alpha}$ in our associated bundle is related to the  connection given in the principal fiber bundle $P(\mathcal M, \mathcal K)$: 
$$B^{\alpha}_i(x,y)={\mathcal A}^{\hat b}_i(x)\, \tilde{\rm e}^{\alpha}_{\hat b}(y),$$
where the mechanical connection ${\mathcal A}^{\hat b}_i(x)$ is obtained as a result of the reduction of the initial dynamical system.
The determinant of our metric 
is given by  $$\det G_{\mathcal A\mathcal B}=\det h_{ij}\,\det {\gamma}_{\alpha\,\beta},\;\;\;\rm{where}\;\;\;\det {\gamma}_{\alpha\,\beta}(x,y)=\det g_{ab}(x)\,[\det\tilde{\rm e}^a_{\alpha}(y)\,]^2.$$

The matrix of the inverse metric $G^{\mathcal A\mathcal B}$ is defined as follows:
\begin{eqnarray*}
\displaystyle
G^{\mathcal A\mathcal B}=\left(
\begin{array}{cc}
h^{ij}  &  - h^{ij}\, B_i^{\alpha}\\
-B_i^{\alpha}\, h^{ij}
& {\gamma }^{\alpha\,\beta}+h^{ij} B_i^{\alpha}B_j^{\beta} 
\end{array}
\right).
\end{eqnarray*}

 The metric $G_{\mathcal A\mathcal B}$ which is under the discussion, is the same  one, up to notation,  that was obtained in \cite{Coquereaux_2, Cho}, and may be in some other papers. For example, the correspondence with the results from \cite{Cho}   is achived by the following identification. 

The projection operator $h^B_A$ \footnote{In our paper we will denote this projection operator by ${\pi}^B_A$.}
 from $\mathscr G$ to $\mathscr G /\mathscr H$  can be presented in the following form: $h^B_A=D^a_A(L_y) \bar{D}^B_a(L_y)$, where $\bar{D}^B_a$ is determined by the equality $\bar{D}^A_b D_A^a={\delta}^a_b$.

In the cited work it was also introduced the ``dual one-form'' $\phi={\phi}^A_{\alpha}dy^{\alpha}$ of the Killing vector filds $K_A$. This form is such that
$$K_A^{\alpha}\,{\phi}^{A}_{\beta}={\delta}^{\alpha}_{\beta},\;\;\;K_A^{\alpha}\,{\phi}^{B}_{\alpha}={\pi}^B_A.$$ 
It can be shown that  ${\phi}^A_{\alpha}=\bar{D}^A_{ b}(L_y)\tilde{\rm{e}}^b_{\alpha}(y)$.
 
We also have the representation for the connection:  $B^A_i(x,y)={\mathcal A}^{\hat b}_i(x)\bar{D}^A_{\hat b}(L_y)$.
Thenб using the representation of these quontities, it is not difficult to establish the coincidens of  our metric (\ref{a1})  with the metric from \cite{Cho}.

\section{Path integral on $\mathcal E$}
The path integrals which we will use in our paper represent the solutions of 
the backward Kolmogorov equations on Riemannian compact manifolds. For the original manifold $\mathcal E$ this equation can be written in the following form:
\begin{equation}
\left\{
\begin{array}{l}
\displaystyle
\left(
\frac \partial {\partial t_a}+\frac 12\mu ^2\kappa \triangle _{\mathcal E}
(q_a)+\frac
1{\mu ^2\kappa m}V(q_a)\right)\psi (q_a,t_a)=0,\\
\psi (q_b,t_b)=\varphi _0(q_b),
\qquad\qquad\qquad\qquad\qquad (t_{b}>t_{a}),
\end{array}\right.
\label{eq3}
\end{equation}
$\mu ^2=\frac \hbar m$ , $\kappa $ is a real positive parameter, $V(q)$ is a $\mathcal G$-invariant potential,
\[
\triangle _{\mathcal E}(q)=G^{-1/2}\frac \partial {\partial q^{\mathcal A}}G^{\mathcal A\mathcal B}G^{1/2}\frac
\partial {\partial q^{\mathcal B}}
\]
is the Laplace--Beltrami operator on 
$\mathcal E$,
$G=\det G_{\mathcal A\mathcal B}$. 

Note that we adopt the convention  by which in our formulas   there is always  a summation over  repeated indices. 
In the previous formula for $\triangle _{\mathcal E}$, the indices denoted by capital letters  run from 1 to $n_{\mathcal E}=\dim {\mathcal E}$.

We assume that the coefficients and the initial function of
(\ref{eq3}) are properly bounded and satisfy the necessary smooth
requirement, so according to \cite{Daletskii}, the solution of the  equation (\ref{eq3})
 can be presented in the  following form:
\begin{eqnarray}
{\psi}_{t_b} (q_a,t_a)&=&{\rm E}\Bigl[\phi _0(\eta (t_b))\exp \{\frac
1{\mu
^2\kappa m}\int_{t_a}^{t_b}V(\eta (u))du\}\Bigr]\nonumber\\
&=&\int_{\Omega _{-}}d\mu ^\eta (\omega )\phi _0(\eta (t_b))\exp
\{\ldots 
\},
\label{eq4}
\end{eqnarray}
where ${\eta}(t)$ is a stochastic process on a manifold 
$\mathcal E$, ${\mu}^{\eta}$ is the path integral measure  on
the path space $\Omega _{-}=\{\omega (t):\omega (t_a)=0, \eta
(t)=q_a+\omega (t)\}$ generated by the probability distribution of a stochastic process $\eta$.

 The global semigroup (\ref{eq4}) is determined by the
path integral which
acts in the space of the smooth and bounded function on $\mathcal E$. This
semigroup is a limit (under the refinement of the subdivision of the time interval) of the local semigroups:
\begin{equation}
\psi _{t_b}(q_a,t_a)=U(t_b,t_a)\phi _0(q_a)=
{\lim}_q {\tilde U}_{\eta}(t_a,t_1)\cdot\dots\cdot {\tilde
U}_{\eta}(t_{n-1},t_b) 
\phi _0(q_a).
\label{eq5}
\end{equation}
Each local semigroup ${\tilde U}_{\eta}$ is built by using a
stochastic family of  local evolution mappings of the manifold $\mathcal E$. 
As for the local semigroups, they
are completely determined by the path integral with the integration measure defined by the local representative ${\eta} ^{\mathcal A}_t$ of the global stochastic process $\eta _t$.  

The local process ${\eta} ^{\mathcal A}_t$ is given by the solution of the stochastic differential equation:
\begin{equation}
d\eta ^{\mathcal A}_t=\frac12\mu ^2\kappa G^{-1/2}\frac \partial {\partial
Q^B}(G^{1/2}G^{\mathcal A\mathcal B})dt+\mu \sqrt{\kappa }{\mathcal X}_{\acute{M}}^{\mathcal A}({\eta}_t)dw^{\acute{M}}_t,
\label{eq6}
\end{equation}
where  the matrix ${\mathcal X}_{\acute{M}}^{\mathcal A}$ are defined by the local equality 
$\sum^{n_{\mathcal A}}_{\acute{\scriptscriptstyle K}
\scriptscriptstyle =1}{\mathcal X}_{\acute{K}}^{\mathcal A}{\mathcal X}_
{\acute{K}}^{\mathcal B}=G^{\mathcal A\mathcal B}$.
(We  denote the Eucledean indices by acute indices.)

The equation (\ref{eq6}) is a Stratonovch equation and it transforms in a covariant way under the changing the chart of the manifold.
This defining property gives one an opportunity to constructa global process on the whole  manifold.

We note that local semigroup ${\tilde U}_{\eta}$ associated with the process ${\eta} ^{\mathcal A}_t$ is defined  by the equality:
\begin{equation}
{\tilde U}_{\eta}(s,t) \varphi (q)={\rm E}_{s,q}\varphi (\eta (t)),\,\,\,
s\leq t,\,\,\,\eta (s)=q.
\label{eq7}
\end{equation}

Therefore, in this definition of the path integral, the transformation of the measure in the path integral which determine the global semigroup is associated the transformation of the local stochastic
differential equations.   

In each chart, the transition from the manifold $\mathcal E$ to the associated bundle $\hat \mathcal E$ means, in our case, that we perform the change of the coordinates $q^{\mathcal A}$ for the coordinates $(x^i,y^{\alpha})$. This transformation leads to the transformation of the stochastic process ${\eta}_t$ for some global process ${\zeta}_t$ generated by the local processes $(x^i_t,y^{\alpha}_t)$. These processes are obtained from the process $q^{\mathcal A}_t$. The transformation of the local processes is in essence the phase-space transformation of the stochastic processes. From the stochastic process theory it is known
that phase-space transformation of the stochastic processes does not
change both the probabilities and the transition probabilities either. 

Taking this into account, we change the local semigroups    
${\tilde U}_{\eta}$,  standing in (\ref{eq5}), for a new local semigroup ${\tilde U}_{\zeta}$.  Therefore we come to a new representation of our global semigroup (\ref{eq5}):
 \begin{equation}
\psi _{t_b}(p_a,t_a)=
{\lim}_q {\tilde U}_{{\zeta}^{\hat\mathcal E}}(t_a,t_1)\cdot\dots\cdot
{\tilde U}_{{\zeta}^{\hat\mathcal E}}
(t_{n-1},t_b) 
{\tilde \varphi} _0(x_a, y_a),
\label{eq8}
\end{equation}
where by ${\tilde U}_{{\zeta}^{\hat\mathcal E}}$ we denote 
\begin{equation}
{\tilde U}_{{\zeta}^{\hat\mathcal E}}(s,t) 
{\tilde \varphi} (x_0,y_0)={\rm E}_
{s,(x_0,y_0)}
\tilde{\phi}(x_t,y_t),
\,\,\,x(s)=x_0,\,\,\,y(s)=y_0.
\label{eq9}
\end{equation}

The equation (\ref{eq8}), representing the path integral on a whole manifold, can be written in the
following symbolical form: 
\begin{equation}
{\psi}_{t_b} (q_a,t_a)={\rm E}\Bigl[\tilde{\varphi
}_0({\xi}(t_b),y(t_b))\exp
\{\frac 1{\mu ^2\kappa m}\int_{t_a}^{t_b}\tilde{V}(\xi
(u))du\}\Bigr],
\label{eq10}
\end{equation}
where $\xi (t_a)=x_a$, $y(t_a)=y_a$, and the boundary values of the processes  are related by the   transformation which performs the transition from the process $\eta _t$ to $\zeta _t$:
$f(q_a)=(x_a,y_a)$.

In (\ref{eq10}), the measure in the path integral  is generated by the stochastic process $\zeta ^{{\mathcal A}}_t=(\xi _t,y_t)$.   The local stochastic differential equations of this process are
\begin{eqnarray}
&&dx^i_t=\frac 12\mu ^2\kappa \Bigl[\frac 1{\sqrt{h{\gamma
}}}\frac
\partial {\partial x^n}(h^{ni}\sqrt{h{\gamma }})
-\frac{h^{ij}}{\sqrt{{\gamma}}}\frac
{\partial} {\partial y^{\alpha}}(B_j^{\alpha}{\sqrt{{\gamma}}})\Bigr]
dt+\mu
\sqrt{%
\kappa }X_{\acute{n}}^idw^{\acute{n}}_t,
\nonumber\\
&&dy^{\alpha}_t=\frac12\mu ^2\kappa \Bigl\{-\frac
{1}{\sqrt{h{\gamma }}%
}\frac {\partial} {\partial x^i}\left( \sqrt{h{\gamma
}}h^{ij}B_j^{\alpha}\right)
+\frac{1}{\sqrt{{\gamma}}}\frac{\partial}{\partial y^{\beta}}\Bigl[
({\gamma }^{\alpha \beta}
+h^{ij}B_i^{\alpha} B_j^{\beta}) \sqrt{{\gamma}}
\Bigr]\Bigr\}dt
\nonumber\\
&&+\mu \sqrt{\kappa }{Y}_{\acute b }^{\alpha} dw^{\acute{b }}_t-\mu \sqrt{\kappa }
X_{\acute{n}}^iB_i^{\alpha}dw^{\acute{n}}_t.
\label{eq11a}
\end{eqnarray}
Here  ${\gamma }=\det \gamma _{\alpha \beta }$,
$h=\det h_{ij}(x)$,     
 $X_{\acute{n}}^i$ and ${Y}_{\acute a}^{\alpha} $
    are defined by the local equalities:
$$\sum_{\acute{n}=1}^{n_{\mathrm{\mathcal M}}}X_{\acute{n}}^i(x)X_{\acute{n}%
}^j(x)=h^{ij}(x)\,\,\, {\rm and} \sum_{\acute{b}=1}^{n_{\mathrm{\mathcal G\backslash\mathcal H
}}}{Y}_{\acute{b }%
}^{\alpha}{Y}_{\acute{b}}^{\beta}={\gamma }%
^{\alpha\,\beta}.$$

In the following we omit the potential term in the path integral as not significant for our transformation.
In this case we can say that 
the differential generator of the semigroup (\ref{eq10}), associated
with the process $\zeta (t)$, is the Laplace-Beltrami operator on the manifold $\mathcal E$, on which the metric, the metric (\ref{a1}),    is obtained from the original metric given on  $\mathcal E$ after transition  to  
the coordinates $(x^i,y^{\alpha})$ of the associated bundle $\hat\mathscr E$.
This operator acts on the scalar functions given on $\mathcal E$. The scalar product in the space of these function is given by the ordinary integral with the standard volume measure obtained from the volume measure $d v_{\mathcal E}(q)=\sqrt{\det G_{\mathcal A\mathcal B}}\,dq^1\cdot\ldots \cdot dq^{n_{\mathcal E}}$ after transition to  new coordinates.
 
 
We note that the stochastic differential equations (\ref{eq11a}) may be simplified. 
It can be shown that the second term of the drift in the stochastic differential equation for  $x^i_t$ is zero. 
This term can be written as  
\[
 {\sqrt{{\gamma}}}\frac
{\partial} {\partial y^{\alpha}}(B_j^{\alpha}{\sqrt{{\gamma}}})=B^{\beta}_i f^{\;\;\;\;\;A}_{DC}{\phi}^D_{\beta}{\phi}^C_{\alpha}K^{\alpha}_A=B^{\beta}_i f^{\;\;\;\;\;A}_{DC}{\phi}^D_{\beta}{\pi}^C_A.
\]
But the terms standing at the right-hand side of this equality is equal to zero for the unimodular group.

The second term of the drift in the stochastic differential equation for $y^{\alpha}_t$ can be transformed as follows
\[
 \frac{1}{\sqrt{{\gamma}}}\frac{\partial}{\partial y^{\beta}}\Bigl[
({\gamma }^{\alpha \beta}
+h^{ij}B_i^{\alpha} B_j^{\beta}) \sqrt{{\gamma}}
\Bigr]=-({\gamma }^{\mu \nu}
+h^{ij}B_i^{\mu} B_j^{\nu})F^{\alpha}_{\mu \nu},
\]
where 
$$F^{\alpha}_{\mu \nu}={\phi}^A_{\mu}F^{\alpha}_{A \nu}=-{\phi}^A_{\mu}\,{\partial}_{\nu}K^{\alpha}_A.$$

The above transformations  represent  the local stochastic differential equations of the process ${\zeta}_t^{\mathcal A}$ in a new form:
\begin{eqnarray}
&&dx^i_t=\frac 12\mu ^2\kappa \Bigl[\frac 1{\sqrt{h{\gamma
}}}\frac
\partial {\partial x^n}(h^{ni}\sqrt{h{\gamma }})
\Bigr]
dt+\mu
\sqrt{%
\kappa }X_{\acute{n}}^idw^{\acute{n}}_t,
\nonumber\\
 &&dy^{\alpha}_t=\frac12\mu ^2\kappa \Bigl\{-\frac
{1}{\sqrt{h{g }}%
}\frac {\partial} {\partial x^i}\left( \sqrt{h{g
}}h^{ij}{\mathcal A}_j^{\hat b}\right)\tilde{\rm e}^{\alpha}_{\hat b}
-
\Bigl(
{\gamma }^{\mu \nu}
+h^{ij}{\mathcal A}_i^{\hat a} {\mathcal A}_j^{\hat b} 
\tilde{\rm e}^{\mu}_{\hat a}\tilde{\rm e}^{\nu}_{\hat b}\Bigr)F^{\alpha}_{\mu\nu} \Bigr\}dt
\nonumber\\
&&+\mu \sqrt{\kappa }{Y}_{\acute b }^{\alpha} dw^{\acute{b }}_t-\mu \sqrt{\kappa }
X_{\acute{n}}^i{\mathcal A}_i^{\hat a}{\rm e}^{\alpha}_{\hat a}dw^{\acute{n}}_t.
\label{eq11}
\end{eqnarray}
In the next section these equations will be used in the derivation of the stochastic nonliner filtering equation. 

\section{Factorization of the path integral measure}
To obtain the transformation of the path integral (\ref{eq9}) which leads to the factorization of the path integral measure  we apply the method from our papers \cite{Storchak}. This method is based on special transformation of the local semigroups which are used in determination of the global semigroup given on the whole manifold.  

The first what one should done, by making use the properties of conditional expectation of the Markov process, is to present the local semigroup as follows:
\begin{equation}
{\tilde U}_{{\zeta}^{\mathcal E}}(s,t) 
{\tilde \varphi} (x_0,y_0)=
{\rm E}
\Bigl[{\rm E}\bigl[\tilde{\varphi }(x(t),y(t))\mid (
{\cal F}_x)_{s}^{t}\bigr]\Bigr],
\label{eq12}
\end{equation}
where the path integral ${\rm E}\bigl[\ldots \mid ({\cal F}_x)_{s}^{t}\bigr]$ is the conditional expectation of a function $\tilde{\varphi }(x(t),y(t))$ given a sub-$\sigma$-algebra generated by the process $x(u),\, u\leq t$.

From the stochastic process theory it is known that in some cases for this conditional expectation there is a nonlinear stochastic equation. The existence of the equation depends on the form of the stochastic differential equations for those  processes that inserted instead of   the  arguments of the function under the expectation. 

If these equation are as follows\cite{Pugachev, Lipster}: 
\begin{eqnarray*}
\left\{
\begin{array}{l}
\displaystyle
dY_t={\varphi}_1(Y_t,Z_t,t)dt+{\psi}_1^{''}(Y_t,t)dW_2(t)\\
dZ_t={\varphi}(Y_t,Z_t,t)dt+{\psi}'(Y_t,Z_t,t)dW_1(t)+\psi''(Y_t,Z_t,t)dW_2(t)
\end{array}\right.
\end{eqnarray*}
then the the equation for $\hat{f}(t)={\rm E}[f(Z_t,t)|Y^t_{t_0}]$, a nonlinear filtering equation, is 
\begin{eqnarray*}
&&d\hat{f}={\rm E}[f_t+f_z\varphi+\frac12f_{zz}(\psi \nu \psi^{\top})(Y_t,Z_t,t)|Y^t_{t_0}]dt
\nonumber\\
&&\;\;\;\;\;\;+{\rm E}[f(Z_t,t)\{\varphi_1^{\top}-\hat{\varphi}_1^{\top}\}+f_z^{\top}(\psi \nu {\psi}_1^{\top})|Y^t_{t_0}]({\psi}_1 \nu {\psi}_1^{\top})(dY_t-\hat{\varphi}_1dt), 
\end{eqnarray*}
with 
$$\hat{\varphi}_1={\rm E}[{\varphi}_1(Y_t,Z_t,t)|Y^t_{t_0}]=\int {\varphi}_1(Y_t,z,t)p_t(z)dz,$$
in which
$p_t(z)$ is a conditional probability density of $Z_t$ with respect to $Y_t$.

Comparing our stochastic differential equations in  (\ref{eq11}) with the above equations, we find the following correspondences between the diffusion terms:
$$\psi'=({\mu\sqrt{\kappa}})Y^{\alpha}_{\acute a},\,\, \psi''=-({\mu\sqrt{\kappa}})B^{\alpha}_iX^i_{\acute n},\,\, 
\psi\, {\psi}_1=-({\mu^2\kappa})h^{ij}B^{\alpha}_j,$$

$${\psi}' {\psi}' +{\psi}'' {\psi}''=({\mu^2\kappa})({\gamma}^{\alpha \beta}+h^{ij} B^{\alpha}_i B^{\beta}_j),\,\,. {\psi}_1=({\mu\sqrt{\kappa}})X^i_{\acute n}.$$
The correspondence between the drift terms of the equations can be easily established.
From this we come to the following equation for the conditional expectation
$\hat{\varphi}(x(t))={\rm E}[\varphi (x(t),y(t))\,|(\mathcal F_x)^t_{t_a}]$:
\begin{eqnarray}
&&d\hat{\varphi}(x(t))=\nonumber\\
&&{\rm E}\Bigl[(\frac{\partial {\varphi}}{\partial y^{\alpha}} )\Bigl(-\frac12\frac{1}{\sqrt{hg}}\frac{\partial}{\partial x^i}\bigl(h^{ij}\sqrt{hg}{\mathcal{A}}^{\hat b}_j(x)\bigr)\tilde{\rm e}^{\alpha}_{\hat b}-\frac12({\gamma}^{\mu\nu}+h^{ij}B^{\mu}_iB^{\nu}_j)F^{\alpha}_{\mu\nu}\Bigr)\Bigr. 
\nonumber\\
&&\Bigl.+\frac12\Bigl(\frac{{\partial}^2  {\varphi}}{\partial y^{\alpha}\partial y^{\beta}}\Bigr)({\gamma}^{\alpha\beta}+h^{ij}B^{\alpha}_iB^{\beta}_j)|({\mathcal F}_{x})^t_{t_a}\Bigr]dt -{\rm E}\bigl[\bigl(\frac{\partial {\varphi}}{\partial y^{\alpha}}\bigr)B^{\alpha}_i|({\mathcal F}_{x})^t_{t_a}\bigr]X^i_{\acute n}dw^{\acute n}_t\nonumber\\
\label{13}
\end{eqnarray}

Let us transform the drift term of the nonlinear filtering  (\ref{13}). First we consider the transformation of  ${\gamma}^{\mu\nu}\,F^{\alpha}_{\mu\nu}\equiv {\gamma}^{\mu\nu}\,(-{\phi}^A_{\mu}\,{\partial}_{\nu}K^{\alpha}_A)$. 

Replacing ${\phi}^A_{\mu}$ and $K^{\alpha}_A$ by ${\bar D}^A_b\tilde {\rm e}_{\mu}^b$ and $D^a_A\tilde {\rm e}^{\alpha}_a$ 
respectively, and using the equation
\[
{\partial}_{\nu}D^a_A(L_y)=f^{\;\;\;a}_{dq}\,\tilde {\rm e}_{\nu}^d \,D^q_a+f^{\;\;\;a}_{hq}\,\tilde {\rm e}_{\nu}^h \,D^q_A+f^{\;\;\;a}_{dh}\,\tilde {\rm e}_{\nu}^d \,D^h_A,
\]
we come (after matrix multiplications) to
\[
{\gamma}^{\mu\nu}\,(f^{\;\;\;a}_{dq}\,\tilde {\rm e}_{\nu}^d\tilde {\rm e}_{\mu}^q+ f^{\;\;\;a}_{hq}\,\tilde {\rm e}_{\nu}^h {\rm e}_{\mu}^q+ f^{\;\;\;a}_{dh}\,\tilde {\rm e}_{\nu}^d \tilde {\rm e}_{\mu}^b\,(D^h_A{\bar D}^A_b)\,)\tilde {\rm e}^{\alpha}_a
+{\gamma}^{\mu\nu}\,\tilde {\rm e}_{\mu}^a({\partial}_{\nu}\tilde {\rm e}^{\alpha}_a).
\]
By the symmetry reason and since on  the homogeneous space (since $D= D(L_y)$) we have $D^h_A{\bar D}^A_b={\delta}^h_b=0$,  the first term is equal to zero.
  Therefore we have
\[
 {\gamma}^{\mu\nu}\,F^{\alpha}_{\mu\nu}={\gamma}^{\mu\nu}\,\tilde {\rm e}_{\mu}^a({\partial}_{\nu}\tilde {\rm e}^{\alpha}_a)=g^{ab}\tilde {\rm e}^{\nu}_b({\partial}_{\nu}\tilde {\rm e}^{\alpha}_a).
\]
Similarly, 
\[
(h^{ij}B^{\mu}_iB^{\nu}_j)F^{\alpha}_{\mu\nu}=(h^{ij}B^{\mu}_iB^{\nu}_j)\,\tilde {\rm e}_{\mu}^a({\partial}_{\nu}\tilde {\rm e}^{\alpha}_a)=(h^{ij}{\mathcal A}_i^{\hat a} {\mathcal A}_j^{\hat b}) 
\tilde{\rm e}^{\nu}_{\hat b}(\partial_{\nu}\tilde{\rm e}^{\alpha}_{\hat a}). 
\]
(Notice that $\tilde{\rm e}^{\mu}_{\hat a}\tilde{\rm e}^a_{\mu}={\delta}^a_{\hat a}$.)

Also note that
\[
g^{ab}\tilde {\rm e}^{\nu}_b({\partial}_{\nu}\tilde {\rm e}^{\alpha}_a)\Bigl(\frac{\partial {\varphi}}{\partial y^{\alpha}}\Bigr)+
g^{ab}\tilde {\rm e}^{\alpha}_a\tilde {\rm e}^{\beta}_b\Bigl(\frac{{\partial}^2  {\varphi}}{\partial y^{\alpha}\partial y^{\beta}}\Bigr)=g^{ab}\tilde {\rm e}^{\alpha}_a\frac{\partial}{\partial y^{\alpha}}\tilde {\rm e}^{\beta}_b\frac{\partial \varphi}{\partial y^{\beta}}.
\]
Taking the obtained equalities into account, we get
\begin{eqnarray}
&&d\hat{\varphi}(x(t))={\rm E}\Bigl[-\frac12\frac{1}{\sqrt{hg}}\frac{\partial}{\partial x^i}\bigl(h^{ij}\sqrt{hg}{\mathcal{A}}^{\hat b}_j\bigr)\,\tilde{\rm e}^{\alpha}_{\hat b}\frac{\partial {\varphi}}{\partial y^{\alpha}} 
+\frac12g^{\bar a \bar b}\,\tilde {\rm e}^{\mu}_{\bar a}\frac{\partial}{\partial y^{\mu}}
\tilde {\rm e}^{\nu}_{\bar b}\, \frac{\partial {\varphi}}{\partial y^{\nu}}\Bigr. 
\nonumber\\
&&
\Bigl.+\frac12({g}^{\hat a\hat b}+h^{ij}{\mathcal A}^{\hat a}_i{\mathcal A}^{\hat b}_j)\,\tilde{\rm e}^{\alpha}_{\hat a}
\frac{\partial}{\partial y^{\alpha}}\tilde{\rm e}^{\beta}_{\hat b}\frac{\partial {\varphi}}{\partial y^{\beta}}\bigl|({\mathcal F}_{x})^t_{t_a}\Bigr]dt 
\nonumber\\
&&-{\rm E}\Bigl[{\mathcal A}^{\hat b}_i\,\tilde{\rm e}^{\alpha}_{\hat b}\frac{\partial {\varphi}}{\partial y^{\alpha}}\bigl|({\mathcal F}_{x})^t_{t_a}
\Bigr]X^i_{\acute n}dw^{\acute n}_t.
\label{filtr-eq}
\end{eqnarray}
In derivation of this equation,  we have used the following decomposition of the metric ${\gamma}^{\alpha\beta}$:
$${\gamma}^{\alpha\beta}=g^{ab}(x)\tilde {\rm e}^{\alpha}_a(y) \tilde {\rm e}^{\beta}_b(y)=g^{\hat a\hat b}(x)\tilde {\rm e}^{\alpha}_{\hat a}(y) \tilde {\rm e}^{\beta}_{\hat b}(y)+g^{\bar a\bar b}(x)\tilde {\rm e}^{\alpha}_{\bar a}(y) \tilde {\rm e}^{\beta}_{\bar b}(y).$$

Now we apply the Peter-Weyl theorem to  the scalar function $\varphi _0 (x^i,y^{\alpha})$  considered  as a function on the coset manifold 
${\mathcal G\backslash \mathcal H}$. In this case it is known \cite{Camporesi, Wawrzynczyk} that the regular representation on the space 
$L^2(\mathcal G\backslash \mathcal H)$ decomposes into direct sum of 
the spherical representations ${\lambda}\in{\hat G}_{\mathcal H}$ of the group $\mathcal G$ \cite{Camporesi}. Notice that for  the scalar function these representations contain only the singlets of the group $\mathcal H$,  
 and the subspace of $\mathcal H$-invariant vectors in the representation space  of every representation ${\lambda}$
is  one-dimensional.

Thus,
\begin{equation}
\varphi _0(x,y)= \sum_{\scriptstyle{\lambda}\in{\hat G}_H}\sum_{\rho=1}^{{\rho}_{\lambda}}\sum_{i=1}^{d_{\lambda}}c^{\lambda}_{\,i\rho}(x)D^{\lambda}_{\,\rho i}(L_y).
\label{decomposHg}
\end{equation}
In this formula  we take into account the fact that  
the irreducible representation of $\mathcal H$  may appears several times in a given representation of  a group $\mathcal G$.
To distinguish these representations  we have used the supplementary index $\rho$ in the spherical representation $D^{\lambda}$.
Note also that in our case $D^{\lambda}(hg)=D^{\lambda}(g)$.

Using (\ref{decomposHg}),
we find that
\begin{equation}
 {\rm E}\Bigl[\varphi (x(t),y(t))\bigl|({\mathcal F}_{x})^t_{t_a}\Bigr]= \sum_{\scriptstyle{\lambda}\in{\hat G}_H}\sum_{i=1}^{d_{\lambda}}\sum_{\rho=1}^{{\rho}_{\lambda}}c^{\lambda}_{\,i\rho}(x(t))
{\rm E}\Bigl[D^{\lambda}_{\,\rho i}(L_{y(t)})\bigl|({\mathcal F}_{x})^t_{t_a}\Bigr].
\label{meanHg}
\end{equation}

It follows that we can reduce the equation (\ref{filtr-eq}) to the corresponding equation for the conditional mathematical expectation
${\hat D}^{\lambda}_{\,\rho i}(x(t))={\rm E}[D^{\lambda}_{\,\rho i}(L_{y(t)})|({\mathcal F}_{x})^t_{t_a}]$.

To perform this we first consider the partial derivatives (with respect to $y$) of the matrix $D^{\lambda}_{\,pq}(L_y)$ belonging to a general irreducible representation of $\mathcal G$. It can be shown that 
\[
 {\partial}_{\mu}D^{\lambda}_{\,pq}(L_y)={\tilde{\rm e}}^A_{\mu}\,D^{\lambda}_{\,ps}(L_y)({J}_A)^{\lambda}_{\,sq}={\tilde{\rm e}}^a_{\mu}\,D^{\lambda}_{\,ps}(L_y)({J}_a)^{\lambda}_{\,sq}+{\tilde{\rm e}}^h_{\mu}\,D^{\lambda}_{\,ps}(L_y)({J}_h)^{\lambda}_{\,sq},
\]
where  $({J}_A)^{\lambda}_{\,ps}$ are the infinitesimal generators of the representation $D^{\lambda}$:  
$ ({J}_A)^{\lambda}_{\,ps}=\frac{\rm{d}}{\rm{d}t}D^{\lambda}_{\,ps}({\rm e}^{tQ_A})\bigl|_{t=0}.$
The second equality holds since in the sum over  the  index $A=(a,h)$, the generator ${J}_A=({J}_a,{J}_h)$. 
Also, in this formula we imply  summation  over the index $s$.

For the spherical represenations ${\lambda}\in{\hat G}_H$, which will be  used in (\ref{filtr-eq}), the generators ${J}_h$ become equal to zero. 
Therefore, in our case, as it follows from the  obtained differentiation formula, we have 
\[
 {\partial}_{\mu}D^{\lambda}_{\,\rho i}(L_y)={\tilde{\rm e}}^a_{\mu}\,D^{\lambda}_{\,\rho s}(L_y)\,({J}_a)^{\lambda}_{\,si},
\]

Using this formula in (\ref{filtr-eq}), we obtain the following equation:
\begin{eqnarray}
&&d{\hat D}^{\lambda}_{\,\rho m}(x_t)=\Bigl[-\frac12\frac{1}{\sqrt{hg}}\frac{\partial}{\partial x^i}\bigl(h^{ij}\sqrt{hg}{\mathcal{A}}^{\hat b}_j\bigr)\,{\hat D}^{\lambda}_{\,\rho s' }(x_t)({J}_{\hat b})^{\lambda}_{\, s'm}\Bigr.
\nonumber\\
&&
+\frac12({g}^{\hat a\hat b}+h^{ij}{\mathcal A}^{\hat a}_i{\mathcal A}^{\hat b}_j)\,({J}_{\hat a})^{\lambda}_{\,q'' s'}({J}_{\hat b})^{\lambda}_{\,s' m}{\hat D}^{\lambda}_{\,\rho q''}(x_t) 
\nonumber\\
&&+\Bigl.\frac12g^{\bar a \bar b}\,({J}_{\bar b})^{\lambda}_{\,s' m}\,({J}_{\bar a})^{\lambda}_{\,q'' s'}{\hat D}^{\lambda}_{\,\rho q''}(x_t)\Bigr]dt -{\mathcal A}^{\hat b}_i\,({J}_{\hat b})^{\lambda}_{\,s m }\,{\hat D}^{\lambda}_{\,\rho s}(x_t)
\,X^i_{\acute n}dw^{\acute n}_t.
\label{filtr_eqHg}
\end{eqnarray}
We note that this equation is a linear matrix stochastic differential equation. 

It is known \cite{Dalmulti,Stroock} that solution of the equation can be presented as follows:
 \begin{equation}
\hat{D}_{\,\rho m}^\lambda (x(t))={\rm E}\bigl[D_{\,\rho q'}^\lambda (L_{y(s)})\mid ({\cal F}
_x)_{s}^t\bigr]\,(\overrightarrow{\exp })_{\,q'm}^\lambda
(x(t),t,s),
\label{29Hg}
\end{equation}
where 
\begin{eqnarray}
&&(\overrightarrow{\exp })_{\,q'm}^\lambda (x(t),t,s)=
\overrightarrow{\exp }%
\int_{s}^t\Bigl\{{\mu}^2\kappa\Bigl[\frac12g^{\bar a \bar b}(x(u))\,({J}_{\bar a})^{\lambda}_{\,q' s'}({J}_{\bar b})^{\lambda}_{\,s' m}
\Bigr.\Bigr.
\nonumber\\
&&+\frac12{g}^{\hat a\hat b}(x(u))({J}_{\hat a})^{\lambda}_{\,q' s'}({J}_{\hat b})^{\lambda}_{\,s' m}-\Bigl.\Bigl.
\frac12\frac{1}{\sqrt{hg}}\frac{\partial}{\partial x^i}
\bigl(h^{ij}\sqrt{hg}{\mathscr{A}}^{\hat b}_j\bigr)\,({J}_{\hat b})^{\lambda}_{\,q' m }
\Bigr]du
\nonumber\\
&&
-\Bigl.{\mu}{\sqrt{\kappa}}
{\mathcal A}^{\hat b}_i(x(u))\,({J}_{\hat b})^{\lambda}_{\,q' m}\,X^i_{\acute n}(x(u))\,dw^{\acute n}(u)
\Bigr\}
\label{30Hg}
\end{eqnarray}
is the multiplicative stochastic integral.
This integral is defined as a limit of the sequence of  time-ordered exponential
multipliers that have been  
obtained as a result of breaking  the time
interval $(t,s)$. The arrow in $(\overrightarrow{\exp })$ is used to indicate  the order of the
multipliers: The arrow  aimed to the multipliers given at greater times.

Using (\ref{meanHg}) and  
(\ref{29Hg})  
in (\ref{eq12}), we get 
\begin{equation}
{\tilde U}_{{\zeta}^{\mathcal E}}(s,t) {\varphi}_0 (x_0,y_0)
=\sum_{\lambda ,i,\rho,q^{\prime }}{\rm E}
\bigl[
 c_{i\rho}^\lambda (x(t))
(\overrightarrow{\exp })_{\,q^{\prime }i}^\lambda 
(x(t),t,s)\bigr] D_{\,\rho q'}^\lambda (L_{y_0}).
\label{31Hg}
\end{equation}

We note that to obtain this equation 
we have taken into account that 
at the initial moment of time 
\[
{\rm E}\bigl[D_{\rho q'}^\lambda (L_{y(s)})\mid ({\cal F}_x)_{s}^t\bigr]
=D_{\rho q'}^\lambda (L_{y(s)})=D_{{\rho} q'}^\lambda (L_{y_0}).
\]

Notice that the expression standing under the sign of the mathemaical expectation in the 
equation (\ref{31Hg})   depends on  a local stochastic process $x(t)$ given on the base
of our fiber bundle.

Using the superposition of the local semigroups obtained after subdivision of the time interval $[t_a,t_b]$  and  
taking a proper limit, we get the global semigroup which can be written as follows:
\begin{equation}
{\psi}_{t_b}(q_a,t_a)
=\sum_{\lambda ,i,\rho,q^{\prime }}{\rm E}
\bigl[
 c_{i\rho}^\lambda (\xi(t_b))
(\overrightarrow{\exp })_{\,q^{\prime }i}^\lambda 
(\xi(t),t_b,t_a)\bigr] D_{\,\rho q'}^\lambda (L_{y_a0}).
\label{32Hg}
\end{equation}  
$(\xi(t_a)=\pi \circ q_a$). 
The process $\xi (t)$ is a global process on a manifold ${\cal
M}={\mathcal E}/{\mathcal G}$.

The infinitesimal  generator of the
semigroup under the sum in equation (\ref{32Hg}) is 
\begin{eqnarray*}
&&\frac 12\mu ^2\kappa \Bigl\{\Bigl[\triangle _M
+h^{ij}\frac 1{\sqrt{g}%
}\Bigl(\frac {\partial \sqrt{g}}
{\partial x^i}\Bigr)\frac \partial
{\partial x^j}\Bigr](I^\lambda )_{pq}
-2h^{ij}{\mathcal{A}}^{\hat b}_i (J_{\hat b})_{pq}^\lambda
\frac \partial {\partial x^j}\Bigr.
\nonumber\\
&&-\Bigl.\frac 1{\sqrt{hg}}\frac
\partial {\partial x^i}\left( \sqrt{hg}h^{ij}{\mathcal{A}}^{\hat b}_j\right) (J_{\hat b})_{pq}^\lambda 
+g^{\bar a\bar b}(J_{\bar a})_{pq^{\prime }}^\lambda (J_{\bar b})_{q^{\prime }q}^\lambda  \Bigr.
\nonumber\\
&&+ \Bigl. (g^{\hat a \hat b }+h^{ij}{\mathcal{A}}^{\hat a}_i{\mathcal{A}}^{\hat b}_j)(J_{\hat a})_{pq^{\prime }}^\lambda (J_{\hat b})_{q^{\prime }q}^\lambda \Bigr\}.
\nonumber\\
\end{eqnarray*} 
(Here  $(I^\lambda )_{pq}$ is a unity matrix.)

This operator acts in the space of functions with the following scalar product
 in the space of the sections of 
the associated co-vector  bundle:
\begin{equation}
(\psi _n,\psi _m)=\int_{\cal M}\langle \psi _n,\psi _m{\rangle}_
{V^{\ast}_{\lambda}}
 \sqrt{g(x)}dv_{\cal M}(x),
\label{33Hg}
\end{equation}
$dv_{\mathcal M}(x)$ is an invariant volume measure on a manifold ${\cal
M}$; in $x^i$-coordinates it is presented as $dv_{\cal
M}(x)=\sqrt{h(x)}dx^1...dx^{n_{\mathcal M}}$.

It is possible to perform the further transformation of the semigroup under the sum in the right side of equation (\ref{32Hg}) is similar to what  was done in \cite{Storchak}. This transformation allows to get rid of the ``redundent'' term in the stochastic equation for the process $\xi (t)$. As a result of the transformation the wave functions leads to a natural normalization without an additional  factor $\sqrt{g(x)}$ in the volume measure. But this, as was shown in \cite{Storchak_2}, give rise to the appearance of the reduction Jacobian in the path integral measure. The geometrical representation of the Jacobian is related to the formula for the scalar curvature for the Kaluza-klein metric. It can be shown that in our problem there exists an analogous relation.

\end{document}